\documentclass[aps,prl,showpacs,twocolumn,floatfix,superscriptaddress]{revtex4}
\usepackage{graphicx}

\begin{document}

\title{Dynamics and steady states in excitable mobile agent systems} 

\date{\today}

\author{Fernando Peruani}
\affiliation{Max Planck Institute for the Physics of Complex Systems, N\"othnitzer Str. 38, 
01187 Dresden, Germany} 
\affiliation{ZIH, 
Technische Universit\"at Dresden, Zellescher Weg 12, 01069 Dresden, Germany}

\author{Gustavo J. Sibona}
\affiliation{Max Planck Institute for the Physics of Complex Systems, N\"othnitzer Str. 38, 
01187 Dresden, Germany}
\affiliation{CONICET and FaMAF, 
Universidad Nacional de C\'ordoba, C\'ordoba, Argentina}

\begin{abstract}
We study the spreading of excitations in 2D systems of mobile agents where the excitation is transmitted when a quiescent agent keeps contact with an excited one during a non-vanishing time. 
We show that the steady states strongly depend on the spatial agent dynamics.
Moreover, the coupling between exposition time ($\omega$) and agent-agent contact rate (CR) becomes crucial to understand the excitation dynamics, which exhibits three regimes with CR: no excitation for low CR, an excited regime in which the number of quiescent agents (S) is inversely proportional to
CR, and for high CR, a novel third regime, model dependent, where S scales with an exponent $\xi -1$, with $\xi$ being the scaling exponent of $\omega$ with CR.
\end{abstract}


\maketitle

The understanding of information propagation through a system of moving agents is crucial for many applications, ranging from chemical reactions to epidemic spreading~\cite{colizza}. 
A particular example of such a process is the  propagation of an excitation through a system of mobile agents. 
In an excitable system, agents typically have three states, quiescent, excited, and refractory, and pass through them in such a way that once excited become first refractory, and
then quiescent again. 
A classical example of an excitable system is the forest-fire~\cite{bak}.  
In this case, the immobility of the agents is well justified.
Another example of excitable system is a disease in which agents undergo a cycle by which an infected agent becomes first recovered and then susceptible (SIRS dynamics). 
Though in the spreading of diseases the mobility of the agent has been for a long time ignored~\cite{kermack,bailey}, its importance has been recently recognized~\cite{hufnagel,cross05,aparicio07,colizza07,sattenspiel94}.
The first attempt to incorporate interactions among individuals was through the study of disease spreading on complex
networks which typically represent social structures~\cite{ballegooijen, joo, kuperman, vespignani_CN,miramonte,gonzalez2,zhang06,mickens}.
The next step was to incorporate the actual motion of individuals across subpopulation structures where nodes represent containers of individuals such as urban areas or just communities~\cite{hufnagel,cross05,aparicio07,colizza07,sattenspiel94}.
Despite the fact that complex networks can
describe realistic social interactions or traffic networks, these models constitute
frozen pictures where edges do not exhibit any dynamics.
Gonzalez et al. in~\cite{gonzalez1} showed that the degree distribution
and clustering coefficient of empirical social networks can be
described by a systems of off-lattice mobile agents, provided the agent density is appropriately chosen. 
Interestingly, in this model agents are constantly moving and the underlying agent-agent contact network is highly dynamical. 

While agent motion may be central to information/excitation spreading, however, it poses questions
that are not easy to answer.  How does the spatial agent dynamics affect the excitation spreading? 
In a SIRS-dynamics, is the agent-agent contact rate (CR) the key parameter that controls the the epidemic dynamics?

In classical epidemiological models~\cite{kermack, bailey} as well as in previous mobile agent models~\cite{miramonte,gonzalez2,zhang06,mickens},  
it is assumed that the mean number of neighbors $\langle \kappa \rangle$ (directly associated to CR) and the effective infection probability $\lambda$ are independent. Consequently, for either a SIRS or SIS dynamics there are only two regimes with CR, resp. $\langle \kappa \rangle$. 
For low CR the system remains in the {\it epidemic extinction} regime, while for CR above a critical value 
the disease propagates such that the number of susceptible agents is inversely proportional to CR.
\begin{figure}[t]
\centering\resizebox{\columnwidth}{!}{\rotatebox{0}{\includegraphics{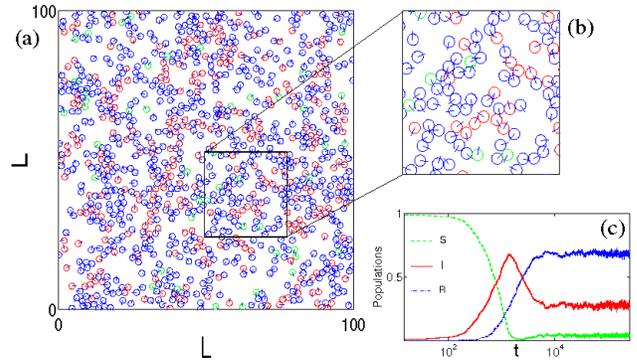}}}
\caption{(a) simulation snapshot. 
Number of agents $N=1024$. 
Colors indicate agent states: quiescent/susceptible (green), excited/infected (red) and
refractory/recovered (blue). 
A short segment indicates the active direction of motion.
(b) amplification of the area indicated in (a). 
(c) temporal evolution of the populations.} 
\label{fig1}
\end{figure}

In this paper we study for the first time an excitable mobile agent system where the excitation is transmitted by keeping physical contact for a finite time (see Fig. \ref{fig1}).
We derive expressions for the threshold and steady states of the excitation dynamics, which highly depend on the spatial agent dynamics,   
and show that the coupling between exposition time (which directly affects $\lambda$) and CR (resp. $\langle \kappa \rangle$) is crucial to understand the evolution of the excitation. 
In particular, we show that due to this coupling the excitation dynamics exhibits three regimes with the CR.
These results applied to epidemic spreading suggests that, contrary to what classical epidemiological models state, 
the epidemic dynamics exhibits more than two regimes with CR. 

{\it Spatial agent dynamics.---}
We assume agents are self-propelled disks which, in absence
of interactions, move at constant speed in a box with periodic boundary condition and change their direction
of motion at Poissonian distributed times.
The equation of motion of the  $i$-th agent can be expressed in
the following way:
\begin{eqnarray}
\label{update_position} \dot{\mathbf{x}}_i(t) =
\frac{\mathbf{F}_i}{\zeta}  + \frac{1}{\zeta} \sum_{j\neq i}
{\mathbf{\nabla}U(\mathbf{x}_i(t),
\mathbf{x}_j(t))}
\end{eqnarray}
where $\mathbf{F}_i/\zeta$ is the active velocity of the
agent and $\zeta$ denotes a friction coefficient. 
As mentioned above, the active direction of motion follows a Poisson process by which an agent changes the direction of $\mathbf{F}_i$ with a rate $\alpha$  while keeping
constant the active speed $\upsilon = \left| \mathbf{F}_i \right| /\zeta$~\cite{param}.

As example of agent-agent interaction we have chosen a repulsive soft-core  
two-body potential that penalizes agent overlapping: 
\begin{equation}
\label{twobody_potential} U(\mathbf{x},
\mathbf{x'}) = \left\{
\begin{array}{ll}
{\gamma} \left[ {|\mathbf{x} - \mathbf{x'}|^{- \beta} - \left(2r\right)^{- \beta}} \right] & \mbox{if $|\mathbf{x} - \mathbf{x'}|<2r$}\\
\\
0 & \mbox{if $|\mathbf{x} - \mathbf{x'}| \geq 2r$}\\
\end{array} \right.
\end{equation} \\
where $r$ is the radius of the agents, $\beta$ is a constant, and $\gamma$ is a function of $\upsilon$ such that the maximum overlapping area between two agents is
independent of $\upsilon$~\cite{param}. 
The collision between two or more agents is a relatively slow
process in which agents keep physical contact for a
non-vanishing time.
Notice that while the implemented dynamics is suitable to describe migration of some micro-organisms~\cite{vicsek_peruani}, it becomes unrealistic as a model for human mobility. 

{\it Excitation dynamics.---}
Without lack of generality we use as excitation dynamics a SIRS dynamics. 
A susceptible  agent gets the disease by keeping
physical contact with an infected one for certain time. 
The transmission of the infection (=excitation) is modeled by a Poissonian process whose characteristic time is $\tau_T$. 
Once infected, an agent carries the infection for a certain period of time, after which  
gets recovered and remains immune to the disease for another period of time, to finally  
become susceptible of being infected.
The transitions from infected to recovered, and from recovered to susceptible are modeled by 
Poisson processes characterized by the characteristic times $\tau_{I}$ and $\tau_{R}$, respectively.
\begin{figure}[t]
\centering\resizebox{\columnwidth}{!}{\rotatebox{0}{\includegraphics{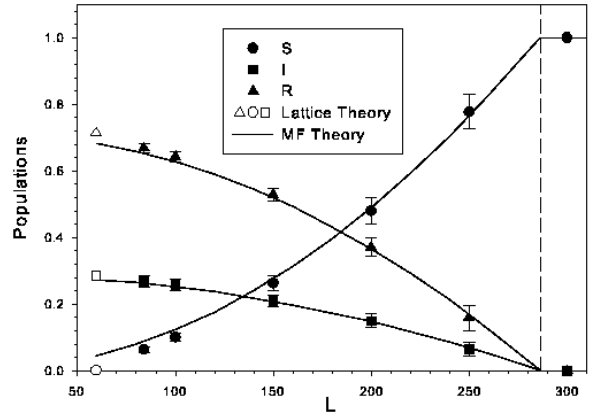}}}
\caption{Populations of agents at the
steady state vs. linear system size $L$. Simulations performed with the parameters given in~\cite{param} and $\upsilon = 0.1$. 
Full black lines corresponds to Eqs. (\ref{s_st_lowdensity}), (\ref{i_st_lowdensity}), and (\ref{r_st_lowdensity})  
while the open symbols to $S^{HD}_{st}$, $I^{HD}_{st}$ and $R^{HD}_{st}$.
The vertical line corresponds to $L_c$.} 
\label{fig2}
\end{figure}

{\it Low density.---}
At  low density  the agents do not form large clusters and
we can reduce the dynamics to a much simpler one in which we
consider only binary collisions. In this gas-like phase we 
describe the time evolution of the populations of susceptible,
infected, and recovered agents by the following mean-field equations:
\begin{eqnarray}
\label{low_s} \dot{S} &=& \frac{R}{\tau_R}  - \left(1-e^{-\omega/\tau_T}\right)
\upsilon \sigma_{0} \rho
 I S \\
\label{low_i} \dot{I} &=& \left(1-e^{-\omega/\tau_T}\right) \upsilon \sigma_0
\rho  IS -
\frac{I}{\tau_I} \\
\label{low_r} \dot{R} &=& \frac{I}{\tau_I} - \frac{R}{\tau_R}
\end{eqnarray}
where $S$, $I$ and $R$ are defined as $S=N_S/N$, $I=N_I/N$ and $R=N_R/N$, 
with  $N$ being the total number of agents in the system, and $N_S$, $N_I$, 
and $N_R$ being the number of susceptible, infected and recovered agents, respectively.
Here, $\sigma_0$ represents the scattering cross
section of the agents, defined as $\sigma_0 = 4r$, $\rho$ is the density ($\rho=N/V$), and $\omega$ denotes the mean duration of the collision
event. 
The exact dependency of $\omega$ with $\upsilon$ depends on the particular agent-agent interaction rule. 
Since in general, for $\upsilon \to \infty$, $\omega$ has to vanish, while for $\upsilon \to 0$, $\omega \to \infty$,  
and considering to the Buckingham $\pi$ theorem~\cite{buckingham}, we conclude that $\omega$ takes the functional form  $\omega\left( \upsilon \right) = K \left( r/
\upsilon \right)^{\xi}$, where $\xi$ is a positive constant and $K$ a dimensional function of the interaction rule parameters. 
We choose values of $K$ and $\xi$ that best fit simulation data of $\omega$ vs. $\upsilon$~\cite{param}.
In consequence, $\left(1-e^{-\omega/\tau_T}\right)$ represents the probability per contact event (of mean duration $\omega$) of transmitting the disease, while $\upsilon \sigma_{0} \rho$ in Eqs. (\ref{low_s}) and (\ref{low_i}) represents the agent-agent contact rate (CR).

From Eqs.~(\ref{low_s})-(\ref{low_r}), 
we obtain the two steady states of
the system: i) $S=1$, $I=0$, and $R=0$, corresponding to the {\it epidemic extinction}, and ii)
\begin{eqnarray}
\label{s_st_lowdensity} S_{st} &=& R_{0}^{-1} \\
\label{i_st_lowdensity} I_{st}&=&\left(1-S_{st}\right)\frac{\tau_I}{\tau_I+\tau_R} \\
\label{r_st_lowdensity} R_{st}&=&\left(1-S_{st}\right)\frac{\tau_R}{\tau_I+\tau_R} \, ,
\end{eqnarray}
\noindent corresponding to the {\it endemic state}, where $R_{0}=\left(1-e^{-\omega/\tau_T}\right)\upsilon \sigma_0 \rho \tau_I$ denotes the {\it basic reproductive number}. 
Recall that $R_{0}>1$ corresponds to  the {\it endemic state}, while $R_{0}<1$ to {\it epidemic extinction}. 
It is instructive to represent $R_{0}$ as $R_{0}=\lambda \langle \kappa \rangle$, where $\langle \kappa \rangle$ denotes the mean number of individuals in contact with an infected agent per time unit, i.e., $\langle \kappa \rangle=\upsilon \sigma_0 \rho \tau_I$, while $\lambda$ refers to the probability for a susceptible individual to contract the disease during a contact event with an infected one, i.e., $\lambda=\left(1-e^{-\omega/\tau_T}\right)$.
Notice that though $\lambda$ and $\langle \kappa \rangle$ are typically independent model parameters, for the current mobile agent-based model, they are coupled. 
This coupling has proved to be essential to understand the disease spreading dynamics.

Fig.~\ref{fig2} shows a comparison of Eqs. (\ref{s_st_lowdensity})-(\ref{r_st_lowdensity}) and agent-based simulations. 
For large values of $L$, there is a good agreement between simulations and theory. 
For small values of $L$, spatial correlations become important and multiple collisions and shielding effects take place.
However, since the spatial dynamics is faster than the excitation dynamics, the mean-field approach still approximates the simulation data as observed in Fig.~\ref{fig2}. 
Clearly, the smaller $L$, the worse the approximation becomes.
Notice that the critical system size $L_c$ that
separates the {\it endemic} from the {\it epidemic extinction} 
regime takes the form: 
$L_c=\left[N \left(1-e^{-\omega/\tau_T}\right) \upsilon \sigma_{0} \tau_I\right]^{1/2}$ 
(dashed vertical line in Fig.~\ref{fig2}).

{\it High density.---}
At high enough densities, agents form percolating clusters and
agent mobility is strongly reduced, though exchange of neighbors still takes place. The maximum conceivable
density, $\rho_M$, corresponds to the one associated to the maximum
packing fraction of disks, $\eta_{M}=\rho_{M} a$, where $a$ is the
area of a single agent.
$\eta_M$ is $0.91$ and corresponds to an hexagonal lattice. 
At  densities close to this limit, 
the system can be reduced to a regular lattice of agents through which
a SIRS dynamics evolves. 
Since we assume that locally there is still some mixing of particles, we neglect correlations between nodes, and express the temporal evolution of agent-populations in terms of a simple mean-field which yields the following {\it endemic} states:
$S^{HD}_{st} = \tau_T/ \left( \tau_I \kappa \right)$, $I^{HD}_{st} = \left(1-S_{st} \right) \left(\tau_I/ \left(\tau_I + \tau_R \right)
\right)$, and $R^{HD}_{st} = \left(1-S_{st} \right) \left(\tau_R/ \left(\tau_I + \tau_R \right) \right)$, where $\kappa$ denotes the mean number of neighbors. 
These values are shown in Fig. \ref{fig2} as open symbols and represent a rough estimate for the system behavior at densities close to $\rho_{M}$.
Better estimates can be obtained by considering spacial correlations through the two-point probability function as indicated in~\cite{joo}. 
Notice that $\kappa$ is constrained to the structure of the lattice and that $S^{HD}_{st} > 0$.
\begin{figure}[t]
\centering\resizebox{\columnwidth}{!}{\rotatebox{0}{\includegraphics{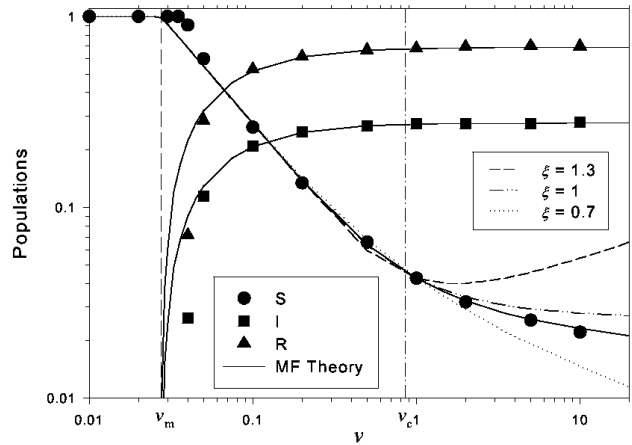}}}
\caption{Populations of agents at the
steady state vs.  agent active speed $\upsilon$. 
Simulations performed with the parameters given in~\cite{param} and $L = 150$.
Full black lines corresponds to Eqs. (\ref{s_st_lowdensity})-(\ref{r_st_lowdensity})).
The vertical lines correspond to $\upsilon_m$ and $\upsilon_c$, while dashed, dotted, and dash-dotted curves correspond to different values of $\xi$, see text.} 
\label{fig_v}
\end{figure}

{\it The role of the active speed.---}
The active speed of the agents $\upsilon$ is a control parameter
which directly regulates the degree of interaction among agents. Particularly, CR is directly proportional to $\upsilon$. 
The values of the {\it epidemic} steady state, as indicated by Eqs. (\ref{s_st_lowdensity})-(\ref{r_st_lowdensity}) and confirmed by agent-based
simulations, exhibit a non-trivial behavior with $\upsilon$ (see Fig. \ref{fig_v}). For low enough density, we expect to observe the
{\it epidemic extinction} state in the limit of $\upsilon$ going to zero.
This implies that to observe the {\it endemic state}, 
$\upsilon$ has to be larger than a threshold value $\upsilon_m$. From Eq.
(\ref{s_st_lowdensity}) we can estimate the minimum speed required by
the disease to survive. This is set by the condition $S_{st}=1$ from
which we get the following transcendental equation for the minimum
speed $\upsilon_m$:
\begin{equation}
\label{trascendental} e^{\frac{- \omega \left( \upsilon_m \right) }{\tau_{T}}} = 1 -
\frac{1}{\upsilon_m \sigma_0 \rho \tau_I} \, .
\end{equation}
By solving numerically Eq. (\ref{trascendental}) for the parameters used in Fig. \ref{fig_v}, we obtain $\upsilon_m = 0.027$  (dashed vertical line).
Note that the expression for $R_{0}$ predicts a crossover at $\omega( \upsilon_c ) = \tau_T$ (dash-dotted vertical line). 

As shown in Fig. \ref{fig_v}, there are three regimes with the agent active speed $\upsilon$. 
For $\upsilon<\upsilon_m$, the mean time between subsequent collisions is larger than the infection period $\tau_I$, and so the disease can not spread and gets extinguished. 
%
On the contrary, for $\upsilon_m<\upsilon<\upsilon_c$, the mean time between subsequent collisions is smaller than $\tau_I$ and the system ends up in an {\it endemic} state.
In this range of $\upsilon$, the mean collision time is such that $\omega(\upsilon) \gg \tau_T$, and thus the probability of transmitting the disease in a collision event is
very high. As a consequence, the epidemic size increases with the CR leading to $S_{st} \sim \upsilon^{-1}$.
Notice that in the limiting case of $\tau_T \to 0$, this regime is valid for all $\upsilon>\upsilon_m$. 
However, any realistic disease has some non-vanishing characteristic transmission time, and consequently, a third regime, ignored so far, emerges for $\upsilon>\upsilon_c$, i.e., for larger values of CR.
In this case, $\omega(\upsilon)<\tau_T$ and the mean collision time is not enough to assure the disease transmission in each collision event. 
The interplay between CR and the transmission probability per collision event determines the third regime which, contrary to the second one, is model dependent and such that $S_{st} \sim \upsilon^{\xi -1}$.
In consequence, there are three possible behaviors depending on the interaction rule, resp. $\xi$. 
If $\xi = 1$, in the limit of $\upsilon \to \infty$ the epidemic size reaches an asymptotic value $S_{\infty} = \tau_T/ \left[K r \sigma_0
\tau_I \rho\right]>0$. 
For $\xi>1$, the disease spreading exhibits a non-monotonic behavior with $\upsilon$ and in the limit of $\upsilon \to \infty$ the systems reaches the {\it epidemic extinction} regime.
Thus, an increase in $\upsilon$, resp. CR, leads to a reduction of the epidemic size. 
Finally, for $0 < \xi < 1$ and in the limit of $\upsilon \to \infty$, $S_{st}$ goes to zero, which indicates that the epidemic size at low density and for large enough values of $\upsilon$ can become larger that the bounded epidemic size corresponding to $S^{HD}_{st}$.
Note that simulations performed with the specific agent-agent interaction  rule given by Eq. (\ref{twobody_potential}) fall into this case, with $\xi = 0.91$.

{\it Concluding remarks.---}
We have shown that in mobile agent systems the coupling between mean exposition time and CR is crucial to understand the excitation dynamics, which generically exhibits three regimes with CR: 
{\it excitation extinction} for low CR, an {\it excited} regime where the excitation is  such  that  
the number of quiescent agents is inversely proportional to CR, and a third regime, for high CR, where the number of quiescent agents scales with an exponent $\xi -1$, with $\xi$ being the scaling exponent of the mean exposition time with CR.
The novel third regime is clearly model dependent and opens, in the context of epidemic spreading, the counter-intuitive possibility of "curing" by increasing CR, which can be achieved in simulations by modifying the interaction rule among agents such that $\xi>1$.

At the experimental level, the introduced mobile agent model might help to understand spatial distribution (and levels) of gene expression induced by cell-cell signaling as observed in some bacterial colonies~\cite{myxo}.
It also might shed some light on disease spreading on bacteria ~\cite{weinbauer98, beretta98} and also on competing strain bacteria experiments~\cite{reichenbach06}.
Clearly, the current agent model is not suitable to describe disease spreading on structured populations, i.e., when the  disease moves across metapopulations, as in humans. 
Further improvements to account for structured communities might include preferred interacting partners, which can be modeled by a long-range potential acting among members of a community, as well as by dividing the space into several connected compartments.

We thank L.G. Morelli for fruitful discussions. 
FP thanks M. B\"ar and A. Deutsch for support. 
We acknowledge funding from SECyT-DAAD (Proalar DA0603).



\begin{thebibliography}{99}



\bibitem{colizza} V. Colizza, R. Pastor-Satorras and A. Vespignani, Nature Physics {\bf 3}, 276 (2007).

\bibitem{bak} P. Bak, K. Chen, and C. Tang, Phys. Lett. A {\bf 147}, 297 (1990).


\bibitem{kermack} W.O. Kermack and G.A. McKendrick, 
Proc. Roy. Soc. A {\bf 115}, 700 (1927).

\bibitem{bailey} N.T.J. Bailey, {\it The mathematical theory of
infectious diseases and its applications. 2nd edition} (Charles Griffin,
Oxford, 1975).


\bibitem{hufnagel} L. Hufnagel, D. Brockmann and T. Geisel, Proc. Natl. Acad. Sci. USA  
{\bf 101}, 15124 (2004).

\bibitem{cross05} P.C. Cross et al., Ecol. Lett. {\bf 8}, 587 (2005).

\bibitem{aparicio07} J.P. Aparicio and M. Pascual, Proc. R. Soc. B {\bf 274}, 505 (2007).

\bibitem{colizza07} V. Colizza and A. Vespignani, Phys. Rev. Lett. {\bf 99}, 148701 (2007).

\bibitem{sattenspiel94} L. Sattenspiel and K. Dietz, Math. Biosci. {\bf 128}, 71 (1995).


\bibitem{miramonte} O. Miramontes and B. Luque, Physica D 168-169, 379
(2002).

\bibitem{gonzalez2}  M.C. Gonzalez and H.J. Herrmann,
 Physica A 340, 741 (2004). M.C. Gonzalez, H.J.
Herrmann and A.D. Araujo, Physica A 356, 100 (2005).

\bibitem{zhang06} D.-M. Zhang et al., Phys. Scr. {\bf 73}, 73 (2006).


\bibitem{mickens} J.W. Mickens abd B.D. Noble, Proceedings of WiSE'05 (2005).


\bibitem{ballegooijen} W.M van Ballegooijen and M.C. Boerlijst, Proc. Natl. Acad. Sci. USA  
{\bf 101}, 18246 (2004).

\bibitem{joo} J. Joo and J.L. Lebowitz, Phys. Rev. E {\bf 70}, 036114
(2004).

\bibitem{kuperman} M. Kuperman and G. Abramson, 
Phys. Rev. Lett. {\bf 86}, 2909 (2001).

\bibitem{vespignani_CN} M. Bogu\~n\'a, R. Pastor-Satorras and A. Vespignani,
Phys. Rev. Lett. {\bf 90}, 028701 (2003);  R. Pastor-Satorras and A. Vespignani, Phys. Rev. Lett. {\bf 86}, 3200 (2001); M. Barth\'elemy et al., J. Theor. Biol.  {\bf 235}, 275 (2005).

\bibitem{gonzalez1} M.C. Gonzalez, P.G. Lind and H.J. Herrmann,
Phys. Rev. Lett. {\bf 96}, 088702 (2006).

\bibitem{param} Simulation parameters. Agent number $N=1024$. Parameters associated with agent motion:
 $\alpha=100$, $\gamma/\zeta= 3.92 \upsilon $, $\beta=1$, and $r=1$. 
For disease dynamics: $\tau_I = 200$, $\tau_R = 500$, and $\tau_T = 1$; 
%
initial fraction of infected agents: $p_{ini}=0.01$.
%
Corresponding to the fitting of $\omega$: $K=1.042$ and $\xi = 0.916$.
%

\bibitem{vicsek_peruani} B. Szab\'o, et al., Phys. Rev E  Phys. Rev. E {\bf 74}, 061908 (2006); F. Peruani, A. Deutsch and M. B\"ar, Phys. Rev E {\bf 74}, 030904(R) (2006).  

\bibitem{buckingham} E. Buckingham, Phys. Rev. {\bf 4}, 345 (1914). 


\bibitem{myxo} M. Dworking and D. Kaiser (Eds.), {\it Myxobacteria II} (Am. Soc. of Microbiology, Washington DC, 1993). 

\bibitem{weinbauer98} M.G. Weinbauer and M.G. H\"ofle, Aquat. Microb. Ecol. {\bf 15}, 103 (1998).

\bibitem{beretta98} E. Beretta and Y. Kuang, Math. Biosci. {\bf 149}, 57 (1998).

\bibitem{reichenbach06} T. Reichenbach, M. Mobilia, and E. Frey, Phys. Rev. E {\bf 74}, 051907 (2006).

\end{thebibliography}
\end{document}